\documentclass{article}
\usepackage{emulateapj,psfig}
\usepackage{graphicx}
\usepackage{timesfonts}
\def\arraystretch{1.25}

\makeatletter
{\end{center}\end{minipage}\smallskip}

\newenvironment{inlinefigure}{%
\def\@captype{figure}%
\noindent\begin{minipage}{0.999\linewidth}\begin{center}}
{\end{center}\end{minipage}\smallskip}
\makeatother
\def\ltsima{$\; \buildrel < \over \sim \;$}
\def\lsim{\lower.5ex\hbox{\ltsima}}
\def\loe{\lower.5ex\hbox{\ltsima}}
\def\gtsima{$\; \buildrel > \over \sim \;$}
\def\gsim{\lower.5ex\hbox{\gtsima}}
\def\goe{\lower.5ex\hbox{\gtsima}}

\newcommand{\be} {\begin{equation}}

\newcommand{\ee} {\end{equation}}

\newcommand{\E}{{\em Einstein}}
\newcommand{\R}{{\em ROSAT}}

\newcommand{\SAX}{{\em Beppo}SAX}
\newcommand{\A}{{\em ASCA}}
\newcommand{\AX}{{\em Chandra}}
\newcommand{\srcsax}{SAX\,J0103.2--7209}
\newcommand{\srcein}{2E\,0101.5--7225}

\newcommand{\bc}{\begin{center}}
\newcommand{\ec}{\end{center}}
\newcommand {\rc}{\rm}

\lefthead{ISRAEL ET AL.}
\righthead{\srcein: A 345\,s PERSISTENT X--RAY PULSAR IN THE SMC}
\submitted{Received: 1999 December 6; Accepted: 2000 January 24}

\begin{document}
\title{\SAX\ and \AX\ Observations of \srcsax\,=\,\srcein: 
a new Persistent 345\,s X--ray Pulsar in the SMC\altaffilmark{\dagger}
}
\authoremail{Gianluca.Israel@oar.mporzio.astro.it}
\author{G.L. Israel\altaffilmark{1,\ast}, 
S. Campana\altaffilmark{2,\ast}, 
S. Covino\altaffilmark{2}, D. Dal Fiume\altaffilmark{3}, 
T.J. Gaetz\altaffilmark{4}, 
S. Mereghetti\altaffilmark{5}, T. Oosterbroek\altaffilmark{6}, 
M. Orlandini\altaffilmark{3}, A.N. Parmar\altaffilmark{6}, 
D. Ricci\altaffilmark{7} and L. Stella\altaffilmark{1,\ast}}

\altaffiltext{0}{$^{\dagger}$The results reported in this Letter are partially 
based on observations carried out at ESO, La Silla, Chile (61.D--0513).}
\altaffiltext{0}{$^{\ast}$Affiliated to the International Center for Relativistic
Astrophysics (I.C.R.A.)}
\affil{1. Osservatorio Astronomico di Roma, Via Frascati 33,
I--00040 Monteporzio Catone (Roma), Italy}
\affil{2. Osservatorio Astronomico di Brera, Via Bianchi 46, I--23807
Merate (Lc), Italy}
\affil{3. TeSRE, C.N.R., 
Via Gobetti 101, I--40129 Bologna, Italy}
\affil{4. Harvard--Smithsonian Center for Astrophysics, 
MS--70, 60 Garden Street, Cambridge, MA 02138, USA}
\affil{5. IFCTR, C.N.R., 
Via Bassini 15, I--20133 Milano, Italy}
\affil{6. Astrophysics Division, Space Science Department of ESA, 
ESTEC, P.O. Box 299, 2200 AG Noordwijk, The Netherlands}
\affil{7. SAX Science Data Center, ASI, Viale Regina Margherita 202, 
I--00198 Roma, Italy}

\begin{abstract}
We report the results of a 1998 July \SAX\ observation of a field in 
the Small Magellanic Cloud (SMC) which led to the discovery of 
$\sim$\,345\,s pulsations in the X--ray flux of 
\srcsax. The \SAX\ X--ray spectrum is well fit by an absorbed 
power--law with photon index $\sim$\,1.0 plus a black body component  
with $kT$\,=\,0.11\,keV. 
The unabsorbed luminosity in the 2--10 keV energy
range is $\sim$\,1.2\,$\times$\,10$^{36}$\,erg\,s$^{-1}$.
In a very recent \AX\ observation 
the 345\,s pulsations are also detected. The available period 
measurements provide a constant period derivative of --1.7\,s\,yr$^{-1}$ over the 
last three years making \srcsax\ one of the most rapidly spinning--up 
X--ray pulsars known.  
The \SAX\ position (30\arcsec\ uncertainty 
radius) is consistent with that of the \E\ source \srcein\  
and the \R\ source RX\,J0103.2--7209. 
This source was detected  at a luminosity level of few 
10$^{35}$--10$^{36}$\,erg\,s$^{-1}$
in all datasets of past X--ray missions since 1979.
The \R\ HRI and \AX\ positions are consistent with that of a m$_{\rm V}$\,=\,14.8 
Be spectral type star already proposed as the likely optical 
counterpart of \srcein. 
We briefly report and discuss photometric and spectroscopic data carried 
out at the ESO telescopes two days before the \SAX\ observation.    
We conclude that \srcsax\ and \srcein\ are the same source, a 
relatively young and persistent X--ray pulsar in the SMC.

\end{abstract}


\section{Introduction} 
During 1997--1998 the number of X--ray pulsars found in the Small 
Magellanic Cloud (SMC) 
rapidly increased from three (SMC\,X--1, RX\,J0053.8--7226, and 
2E\,0050.1--7247; Lucke et al 1976; Hughes 1994; Israel et al. 
1997) up 14 (for a review see Yokogawa et al. 1998) 
thanks to sensitive observations of the large area detectors on 
board the {\em R}XTE and ASCA satellites.
The majority were found to be associated with massive Be  
spectral type stars showing intense H$\alpha$ emission lines.
Only SMC\,X--1, which is associated with 
a super--giant B0 spectral type  star in a 3.9\,d orbital period 
binary system, is a persistent (although moderately variable) X--ray 
pulsar. For all the remaining X--ray pulsars pronounced variability 
(a factor $\geq$\,50) or, more often, transient behaviour 
has been definitively proven.  

The source \srcein\ was detected at a nearly constant flux level 
in all the \E, \R\ and \A\ pointings which surveyed the relevant 
region of the SMC (see Hughes \& Smith 1994),
but pulsations were not found due to poor statistics. 
Based on the accurate position obtained with the \R\ HRI, these authors 
found that  \srcein\ is very likely associated with 
a Be spectral--type star (R.A.\,=\,01$^{\rm h}$03$^{\rm m}$13\fs86, 
Dec.\,=\,--72\arcdeg09\arcmin14\farcs1; equinox J2000). 
\srcein\ is located near the optical limb of the supernova remnant 
SNR\,0101--72.4. Hughes \& Smith (1994) present 
several  arguments that 
make the Be/X--ray binary -- SNR association unlikely.

In this {\it Letter} we report the discovery of 345\,s pulsations from the 
source \srcsax\ during a \SAX\ 
observation of the SMC. The comparison with the data of past X--ray 
missions allows us to conclude that \srcsax\ and \srcein\ are the same object, 
a persistent source 
with  moderate variability (within a factor of 5--10). 
We also report the results of the timing analysis of a recent public  
\AX\ observation and discuss optical observations 
carried out at ESO. 

\section{Observations and Data Analysis}

\subsection{Beppo{\em SAX} observation}
The SMC field including the position of 
the \srcein\ was observed by the Narrow Field 
Instruments (NFIs) on board the \SAX\ satellite (Boella et al. 1997a)
on 1998 July 26--27 (effective exposure time of 40320\,s). 
We used data from the Medium Energy (MECS; Boella et al. 1997b) and Low Energy 
(LECS; Parmar et al. 1997) instruments.   
A bright X--ray source ($\sim$\,3.7\,$\times$\,10$^{-2}$\,ct\,s$^{-1}$, 
1--10\,keV) was detected on--axis in the MECS, 
at R.A.\,=\,01$^{\rm h}$03$^{\rm m}$13$^{\rm s}$, 
Dec.\,=\,--72\arcdeg09\arcmin16\arcsec\ (J2000; 90\% confidence uncertainty  
radius of 30\arcsec\ ).
The MECS event list and 
spectrum were extracted from a circular region of 4\arcmin\ radius 
(corresponding to an encircled energy of $\sim$\,90\%) around the 
X--ray position. 
\begin{figure*}
\centerline{\includegraphics[width=0.25\linewidth]{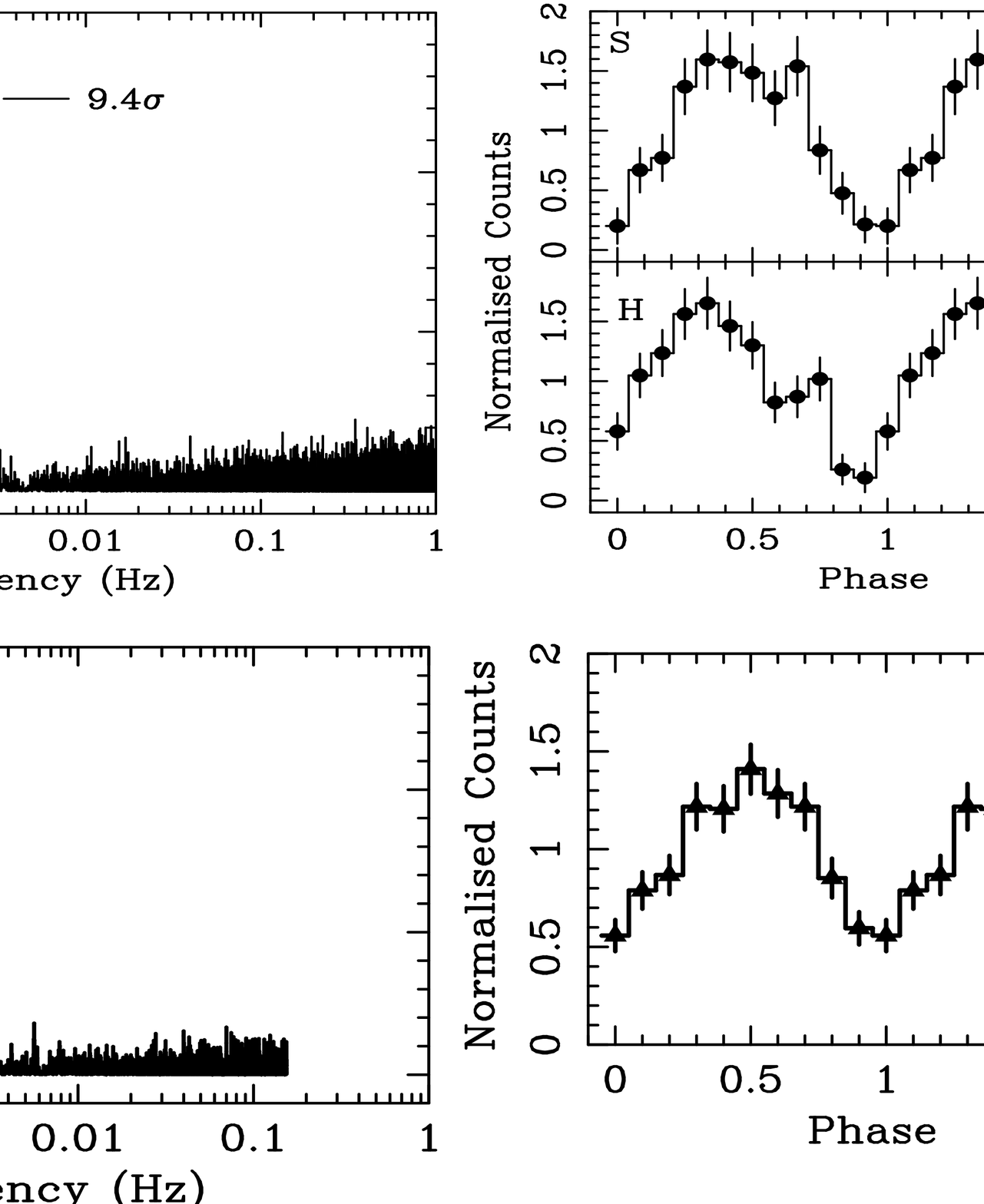}
\includegraphics[width=0.205\linewidth]{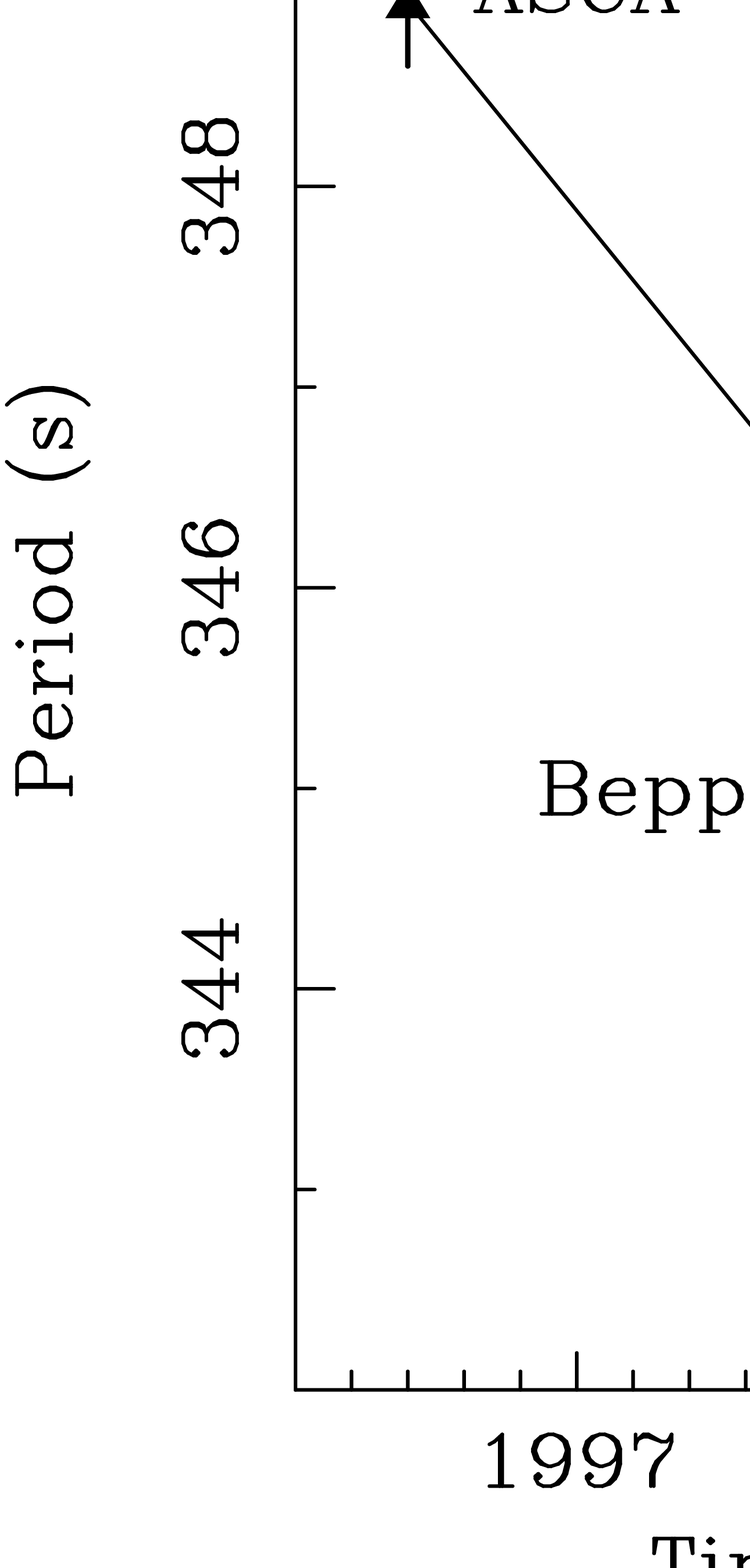}}
\caption{1998 July (upper left panel) and 1999 August (lower left panel) 
power spectra of the 1--10 keV \SAX\ MECS and \AX\ ACIS light curves of 
\srcein. The corresponding folded light curves at the best periods are 
also shown in the central panels. The \SAX\ folded light curve is given 
separately in the 1--4\,keV (S) and 4--10\,keV (H) ranges. The period 
history of \srcein\ is shown in the right panel.}
\end{figure*}
\begin{inlinefigure}
\centerline{\includegraphics[width=0.60\linewidth]{fig2alte.ps}}
\caption{\SAX\ (crosses) and \R\ PSPC (triangles) spectra of 
\srcein. The best fit model is shown, together with the corresponding residuals.}
\end{inlinefigure}
A 4\arcmin\ extraction radius ($\sim$\,85\% encircled energy) 
was also used for the LECS in order to minimize the contamination from 
the soft X--ray emission of the bright nearby source 2E\,0102.3--7217. 
The local background was measured in a region of the 
MECS and LECS images far from any detected field sources.

The arrival times of the $\sim$\,1500 photons were corrected to the 
barycenter of the solar
system and background subtracted light curves were accumulated in 0.5\,s
bins. A single power spectrum was calculated over the entire time span 
covered by the observation 
in order to maximize the sensitivity to coherent pulsations. Significant 
power spectrum peaks were searched for 
using the algorithm described in Israel \& Stella (1996).
A highly significant peak ($\sim$\,9.4\,$\sigma$ based on the fundamental 
only; see upper left panel of Fig.\,1) 
was found at a frequency of 0.002889\,Hz, corresponding to a period of 345.2\,s. 
An accurate determination of the period was obtained by fitting the phases of the 
modulation over 4 different intervals of $\sim$\,16000\,s each. 
The scatter of the phase residuals was consistent with a strictly periodic  
modulation at the best period of 345.2\,$\pm$\,0.3\,s (90\% confidence).
A comparison of the 1--4\,keV (panel S in Fig.\,1) 
and 4--10\,keV (panel H) folded light curves provides marginal evidence 
for an energy dependent  pulse  profile: nearly sinusoidal at low energies, 
while double--horned above 4\,keV. However the pulsed fraction 
(semi--amplitude of modulation divided by the mean 
source count rate) of $\sim$\,45\% is constant over the two energy 
intervals considered. 
  
A spectral analysis was performed in the 0.7--6.5 and 1.6--10\,keV energy 
ranges for the LECS and MECS, respectively. The spectra were rebinned 
to have at least 20 counts in each energy channel, such that 
$\chi^{2}$ fitting techniques could be used. No single component model 
was found to fit the data well (a power--law gave a 
$\chi^2/dof$\,=\,41/15; {\rc where $dof$ is the degree of freedom}). 
Among double component models, an absorbed power--law plus a black 
body gave the best fit ($\chi^2/dof$\,=\,15/15) for a photon index of 
1.0\,$\pm^{0.2}_{0.1}$,  
N$_H$\,$\leq$\,3.8$\pm_{3.8}^{7.5}$\,$\times$\,10$^{21}$\,cm$^{-2}$, and a black 
body temperature of 0.11\,$\pm$\,0.03\,keV (uncertainties refer to 
1$\sigma$).
The observed flux in the 2--10 keV energy band was 
2.7\,$\times$\,10$^{-12}$\,erg\,s$^{-1}$ cm$^{-2}$ (the soft 
component accounts for $\sim$\,15\% of the total) corresponding to 
an unabsorbed 2--10\,keV X--ray luminosity of 
1.2\,$\times$\,10$^{36}$\,erg\,s$^{-1}$ assuming a distance 
of 62\,kpc (Laney \& Stobie 1994). 

\subsection{ Archival X--ray Observations} 
After the discovery of 345\,s pulsations 
in \srcein\ (Israel et al. 1998), Yokogawa \& Koyama (1998) found a 
signal at a period of 348.9\,s during a 1996 May \A\ observation 
(absorbed luminosity of 5\,$\times$\,10$^{35}$\,erg\,s$^{-1}$; 
2--10\,keV) implying a period derivative (with respect to \SAX) of 
--1.7\,s\,yr$^{-1}$. The source was also detected by \A\ on 1993 May 12 
and 1997 November 14 at an absorbed luminosity level of 6 and 
5\,$\times$\,10$^{35}$\,erg\,s$^{-1}$, respectively. A re--analysis 
of the latter \A\ datasets allowed us to infer an upper limit 
(3$\sigma$ confidence) in the 60--80\% range on the pulsed fraction for 
a period of $\sim$\,345\,s.

To better constrain the absorption we used archival data 
from the \R\ satellite (PSPC; 0.1--2\,keV band) which observed the field 
of \srcein\ between 1991 October 8 and 1992 April 28 (sequence 600195; 
effective exposure time 26630\,s). 
Assuming that the source showed the same flux and spectrum during the 
two observations, we fitted the \R\ and \SAX\ data together.
We found that the two component model derived above again gave the best fit;  
a $\chi^2/dof$\,=\,23/25 was 
obtained for a  photon index of 1.1\,$\pm$\,0.1 with N$_H$ of 
3.9$\pm_{2.4}^{4.0}$\,$\times$\,10$^{21}$\,cm$^{-2}$ and a black body 
temperature of 0.12$\pm_{0.03}^{0.04}$\,keV corresponding to an 
equivalent black body radius R$_{bb}$ in the 2--60\,km range (see Fig.\,2; 
1$\sigma$ uncertainties). 
The corresponding power spectrum shows no significant (3$\sigma$ 
confidence) peak in the 339--359\,s period interval  
with a $\sim$\,60\% upper limit on the pulsed fraction.  
\begin{figure*}[thb]
\centerline{\includegraphics[width=0.396\linewidth]{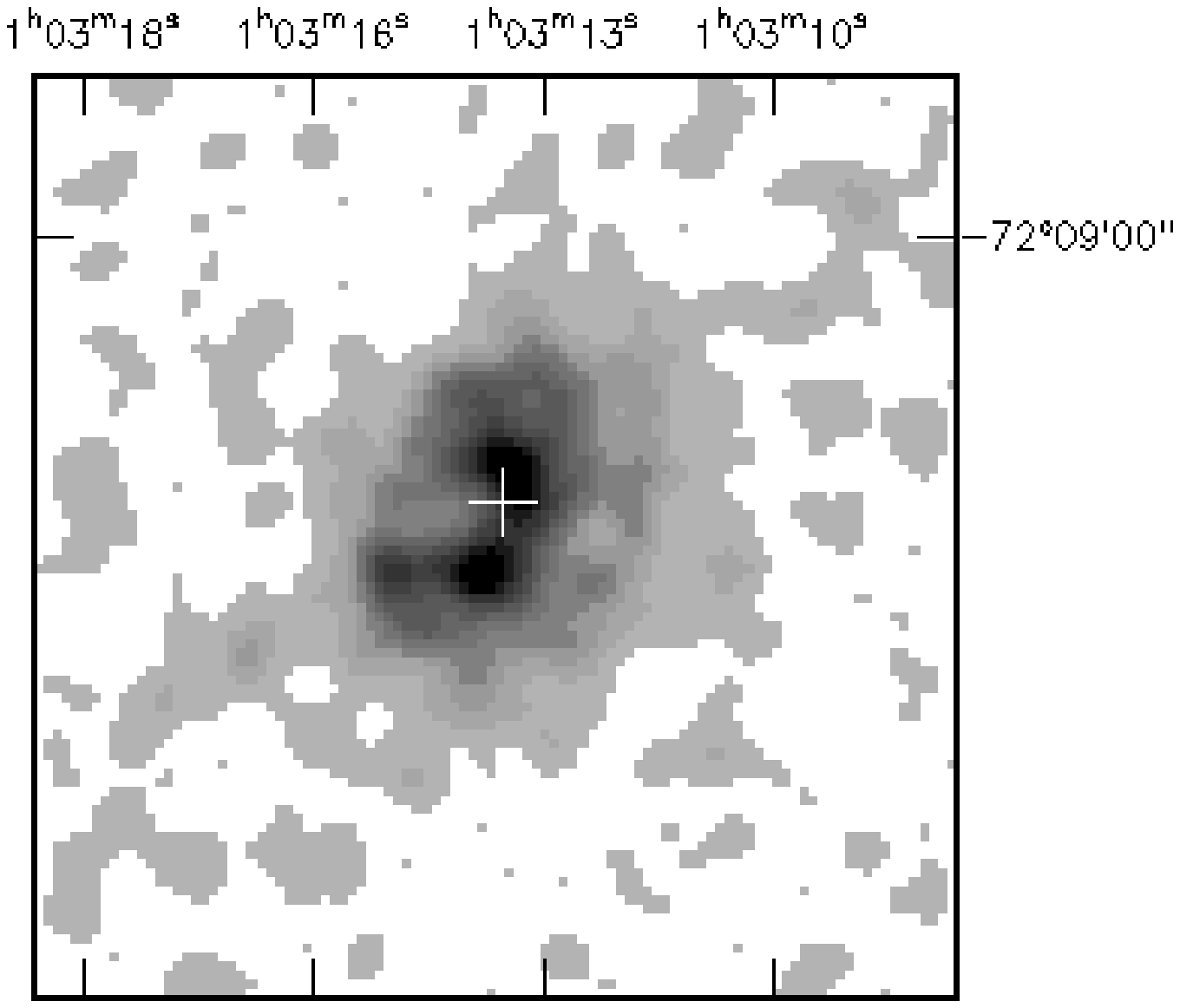}
\includegraphics[width=0.288\linewidth]{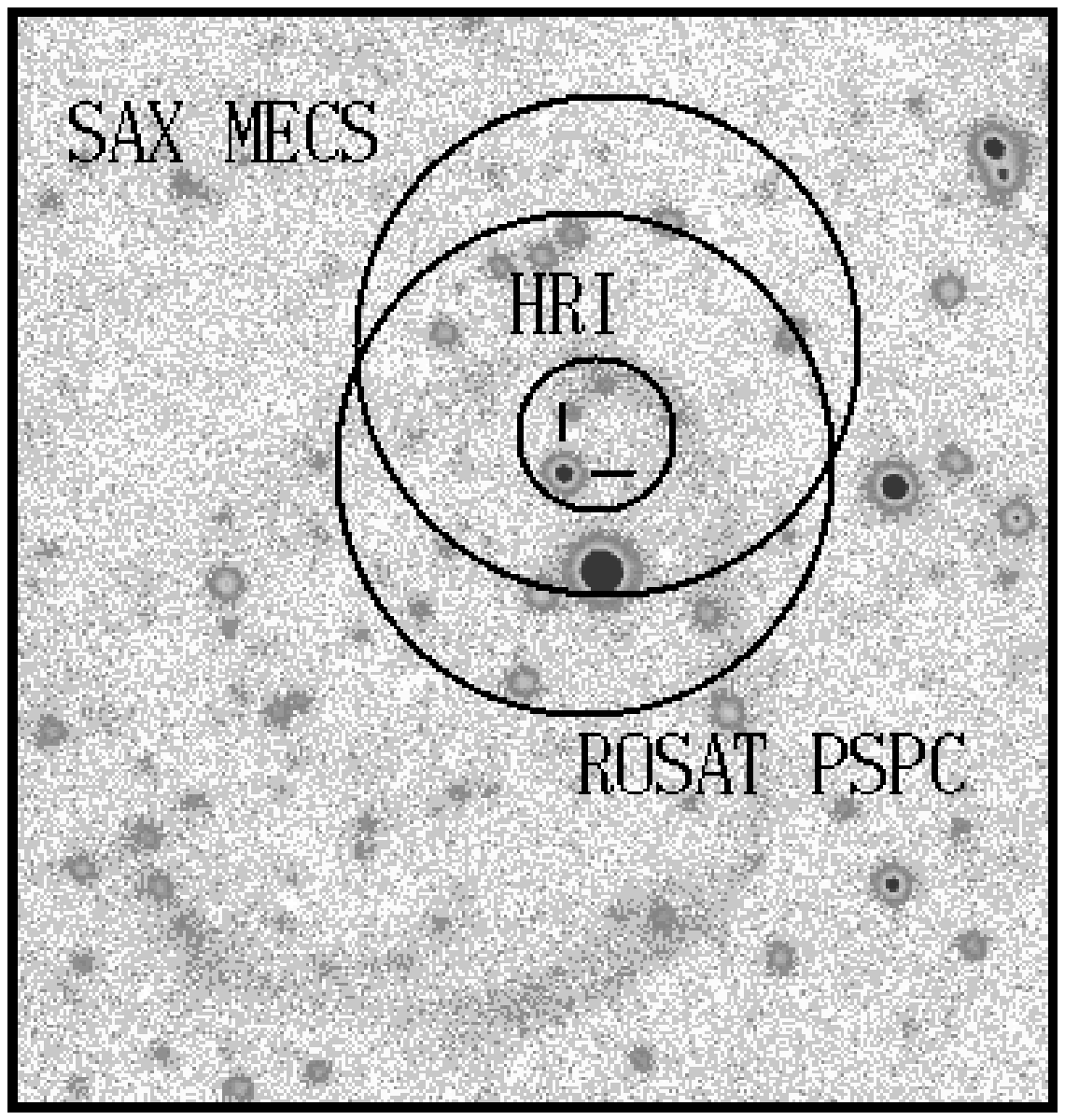}}
\caption{45\arcsec$\times$45\arcsec\ \AX\ ACIS image of the region 
around \srcein\ (left panel). The image was smoothed with a Gaussian of 
$\sigma$=2\arcsec. The cross marks the optical 
position of the Be star identified by Hughes \& Smith (1994). 
The double--peaked shape of the source is artificial (see text for details). 
2\arcmin$\times$2\arcmin\ H$\alpha$ image together with the 
\R\ and \SAX\ position circles (right panel). The proposed optical 
counterpart is also marked.}
\end{figure*}

\subsection{Chandra observation}
The field including \srcein\ was observed 
by the NASA Advanced X--ray Astrophysics Facility satellite \AX\  
on 1999 August 23 with the Imaging Spectrometer (ACIS; 
Garmire et al. 1992, Bautz et al. 1998 and references therein) 
in the high--resolution imaging mode for an effective exposure time 
of 19551\,s. The source was detected at an off--axis angle of about 
9\arcmin\ in the S4 (front illuminated) CCD of the ACIS--S detector
with a count rate of 
$\sim$\,9\,$\times$\,10$^{-2}$ ct s$^{-1}$  (in the 0.5--10\,keV energy 
range; see Fig.\,3 left panel). The source shows two 
emission peaks separated by $\sim$\,5\arcsec. 
However 
the size of the emission region is comparable with 
the 90\% encircled energy region of a source at 9\arcmin\ off--axis angle.  

We extracted a 3.24\,s binned light curve of the source from a region 
within 8\arcsec\  
from the center of the emission: R.A.\,=\,01$^{\rm h}$03$^{\rm m}$14\fs06, 
Dec.\,=\,--72\arcdeg09\arcmin15\farcs25 (equinox J2000). The statistical 
uncertainty 
radius is only $\sim$\,0.5\arcsec\ but at this stage the uncertainty in the 
absolute positioning at such an off--axis angle  
might be as large as 90\arcsec. However 
the detection of the pulsations clearly associates the \AX\ source 
with that of \SAX. 
We also note that the \AX\ coordinates are consistent with the \R\ HRI 
uncertainty circle and 
differ by less than 2\arcsec\ from those of the proposed optical 
counterpart (see below) making the association very likely.   
According to the \AX\ source naming convention we designated it as 
CXO\,J010314.1--720915.

A single power spectrum was calculated over the entire observation. 
The search was performed over a period interval around that detected 
by \SAX\ and assuming a maximum $|$\.P$|$ of $\sim$3 s yr$^{-1}$ 
which translates into a search over 
only two Fourier frequencies (see Fig.\,1 lower panels). A peak was detected 
at a significance level of 7.5$\sigma$. A refined period was 
determined by means of the phase fitting technique; this gave a value of 
343.5\,$\pm$\,0.5\,s (see Table\,1). We note that the \AX\ data were not 
corrected to the barycenter of the solar system. 
However, for a relatively long period pulsar in the direction of the SMC, 
the effect of 
the spacecraft and of the Earth motion, would cause a maximum correction  
of a factor of 10 smaller than the statistical uncertainty of the period given 
above. The pulse profile is sinusoidal 
(over the whole \AX\ energy band) and the pulsed fraction is $\sim$\,45\% 
(see Fig.\,1).  This result implies that the pulsar is continuing to spin--up 
at a constant rate of --1.7\,s\,yr$^{-1}$ since 1996 (see Fig.\,1; right panel).   
  
In order to further address the issue of the double peaked  
\AX\ image (see Fig.\,3), we extracted two separate light 
curves for each of the two peaks. 
We found that the signal at 343.5\,s was present with a similar pulsed fraction
in both light curves; this result is consistent with the hypothesis of 
a point--like source. An independent confirmation was also obtained through a 
raytracing simulation, encompassing a monochromatic source (1.49\,keV) and 
models for the mirror assembly and ACIS detector. The results indicate that 
expected PSF at an off--axis angle of 9\arcsec\ is artificially elongated in 
the direction connecting the two peaks detected in the Chandra image around 
the position of \srcein; the two peaks are probably artefacts resulting 
from the small number of X--rays in the image. 
We conclude that the source is consistent with being 
point--like.\vspace{-2mm}

\subsection{Optical Observations}
Optical images (H$\alpha$ and H$\alpha$ red--continuum; 200\,s each) of the 
\SAX\ error circle of 
\srcsax\ were obtained on 1998 July 24 at the Danish 1.5\,m 
telescope with the Danish Faint Object 
Spectrometer Camera (DFOSC) at La Silla (Chile) in order to search for 
emission--line stars. 
The data were reduced using standard ESO--MIDAS procedures for 
bias subtraction, flat--field correction, aperture photometry and one 
dimensional stellar and sky spectra extraction. 
Within the \SAX\ position circle we found only one H$\alpha$ active object: 
this is the O9--B1\,III--Ve 
m$_{\rm V}$=14.8 star  originally suggested as the optical 
counterpart of \srcein\ (Hughes \& Smith 1994). 
A 10\,\.A   resolution 3800--8500\,\.A spectrum (2\farcs0 slit) of this  
star was obtained with the same instrument on 1998 
October 20. Strong H$\alpha$ and H$\beta$ emission--lines were 
detected, with equivalent width   --20\,$\pm$\,2\,\.A, and 
--1.8\,$\pm$\,0.2\,\.A, respectively. These results are in good agreement 
with those obtained by Hughes \& Smith (1994) during observations 
carried out on December 1992.
\begin{table*}
\begin{center}
\begin{tabular}{ccccccc}
\multicolumn{7}{c}{\small TABLE\,1}\\
\multicolumn{7}{c}{\small \sc Period history for \srcein.}
\\ \hline \hline
\small Mission &\small Instrument     & \small Period  &\small Date &\small Exposure      &\small 2--10\,keV L$_X$ &\small Reference\\
        &               &\small  (s)     &      & \small (s)   &\small  (10$^{35}$\,erg\,s$^{-1})$   &\\

\hline 
\small \A\     &\small  GIS           &\small  348.9\,$\pm$\,0.3   &\small  1996 May 21--23 &\small  32912 &\small  5 &\small  1 \\
\small \SAX\   &\small  MECS          &\small  345.2\,$\pm$\,0.2   &\small  1998 Jul 26--27 & \small 40320 &\small  11 &\small  2; this work \\
\small \AX\    &\small  ACIS          &\small  343.5\,$\pm$\,0.5   &\small  1999 Aug 23     &\small  19551  & --- &\small  this work \\
\hline
\end{tabular}
\begin{flushleft}
\small \noindent Note: Luminosities refer to absorbed fluxes. 90\% 
uncertainties are reported.\\ 
\noindent 1 -- Yokogawa \& Koyama (1998); 2 -- Israel et al. (1998).
\end{flushleft}
\end{center}
\end{table*}

\section{Discussion}
Although \srcein\ shares several characteristics with other  
X--ray pulsators in the Magellanic Clouds (MCs) and in the 
Galactic plane (i.e., 
the spin period and pulsed fraction, the spectral shape 
at high energy and the presence of a soft thermal component, 
the association with a high mass companion, etc.), it has also some 
peculiar differences which make it unusual. 

The absence of long--term variability greater than a factor of $\sim$5--10
over an interval of $\sim$\,20 years is strongly suggestive of a {\em 
persistent} X--ray pulsar, the second in the SMC since the discovery of 
pulsations from  
SMC\,X--1. Moreover, the association of the companion with a Be 
spectral--type star makes \srcein\ the first example of a persistent 
(main sequence) Be/X--ray binary system in the SMC. 
\srcein\ is also a relatively low--luminosity 
(10$^{35}$--10$^{36}$\,erg\,s$^{-1}$) X--ray pulsar.    
Finally, the inferred  period derivative of --1.7\,s\,yr$^{-1}$ 
corresponds to a secular spin--up time--scale of $\sim$\,200\,yr which is the 
shortest of any known X--ray pulsar in a High Mass X--ray binary. 

By using the P and \.P measurements and an 
X--ray luminosity in the 0.5--2\,$\times$\,10$^{36}$ erg 
s$^{-1}$ range, we infer a magnetic field of 
4--12\,$\times$\,10$^{12}$\,Gauss 
and, correspondingly, a maximum magnetospheric radius of $r_{\rm m}$
$\sim$\,4\,$\times$\,10$^9$\,cm (see Lamb et al. 1973). 
Since the derived corotation radius $r_{\rm co}$ is  
$\sim$\,8\,$\times$\,10$^9$\,cm, the $r_{\rm m}$ is considerably  
smaller than $r_{\rm co}$, and accretion on the surface of 
the neutron star proceeds unaffected by the magnetosphere's centrifugal 
drag, as long as the luminosity of the 
source is larger than $\sim$\,2\,$\times$\,10$^{35}$\,erg\,s$^{-1}$. 
Such a luminosity is not unusual for a bright X--ray pulsar in a nearly 
circular orbit around a giant OB companion star where the intense stellar 
wind continuously supplies matter to the compact object. 
Assuming a main sequence star, an orbital period of $\sim$\,300\,d 
is expected for \srcein\ based on the pulse period -- orbital period 
correlation  of Be star/X--ray pulsar binaries (Corbet 1984).
A timing analysis of I band photometric measurements from the Optical 
Gravitational Lensing Experiment (OGLE) revealed no evidence of any 
coherent signal in the 1--50\,d period interval (see Coe \& Orosz 1999) 
suggesting that \srcein\ is a long period system.

The characteristics of \srcein\ have their closest analogy to those of 
the persistent 
low--luminosity (10$^{35}$--10$^{36}$\,erg\,s$^{-1}$; 2--10\,keV band) 
X--ray pulsar RX\,J0146.9+6121, a 1455\,s (spin--up timescale of 
$\sim$\,250 yr) spinning neutron star in a binary system with a B1Ve 
companion star (see Reig et al. 1997; Haberl et al. 1998; Mereghetti et al. 
2000).  Two other recently discovered long period Be/X--ray 
pulsars in the SMC, namely AX\,J0051--733 (P$_s$=323\,s; Imanishi 
et al. 1999) and AX\,J0058--7203 (P$_s$=280\,s; Tsujimoto 1999), although 
highly variable (a factor of 10--100), have a maximum luminosity level 
close to that of \srcein. 
             
In conclusion, we discovered a new X--ray pulsar with a period of 
$\sim$\,345\,s in the SMC which is persistent, has a relatively 
low--luminosity, is rapidly spinning--up and is associated with a 
O9--B1\,III--Ve star. The relatively high value inferred for the magnetic 
field and period derivative point to a young object. 
All these findings  make \srcein\ an unusual X--ray pulsar  
which deserves more detailed investigations with instrumentation 
ranging from the X--ray to the IR band.  

\acknowledgments
This work was partially supported through ASI and CNAA grants.
We thank the \AX\ Data Archive (CDA) of the \AX\ X--Ray Observatory Science 
Center (CXC) at CfA for a prompt release of the data.  
We thank R. Ragazzoni for obtaining the optical spectrum of the 
counterpart of \srcein. We thank H. Tananbaum, R.J. Edgar and L. Burderi 
for their helpful comments. 
This work has been supported in part by NASA contract NAS8-39073.

\end{document}